\documentclass[pre,twocolumn,superscriptaddress,amsmath,amssymb,showpacs,noshowkeys,a4paper]{revtex4}
\usepackage{amsmath}
\usepackage{amssymb}
\usepackage{amsfonts}
\usepackage{graphicx}
\usepackage{dcolumn}
\usepackage{bm}
\usepackage[usenames]{color}
\usepackage{epstopdf}
\definecolor{mygray}{gray}{0.5}

\def\pdiff#1#2{{{\partial #1} \over {\partial #2}}}

\newcommand{\afflangevin}{\address{Institut Langevin, ESPCI Paristech, CNRS - UMR 7587, PSL Research University, Universit\'e Pierre and Marie Curie,  1 Rue Jussieu, 75005, Paris, France, EU}}
\newcommand{\affMSC}{\address{Laboratoire Mati\`{e}re et Syst\`{e}mes Complexes, Universit\'e Paris Diderot, Sorbonne Paris Cit\'{e}, CNRS - UMR 7057, 10 Rue A. Domon and L. Duquet, 75013 Paris, France, EU}}
\newcommand{\affMPQ}{\address{\emph{Present address:} Laboratoire Mat\'{e}riaux et Ph\'{e}nom\`{e}nes Quantiques, Universit\'e Paris Diderot, Sorbonne Paris Cit\'{e}, CNRS UMR 7162, 10 Rue A. Domon and L. Duquet, 75013 Paris, France, EU}}
\newcommand{\affJFI}{\address{\emph{Present address:} Department of Physics and James Franck Institute, University of Chicago, Chicago, 929 E 57th St, IL 60637, USA}}
\newcommand{\affMIT}{\address{Department of Mathematics, Massachusetts Institute of Technology, 77 Massachusetts Avenue, Cambridge, MA 02139, USA}}
\newcommand{\affNYU}{\address{Courant Institute of Mathematical Sciences, New York University, New York, NY 10012, USA}}

\begin{document}
\title{Pilot-wave dynamics in a harmonic potential:\\ Quantization and stability of circular orbits}
\author{M. Labousse}
\afflangevin
\affMSC
\affMPQ
\author{A.~U. Oza}
\affNYU
\author{S. Perrard}
\affMSC
\affJFI
\author{J.~W. M. Bush}
\email{bush@math.mit.edu}
\affMIT
\begin{abstract}
We present the results of a theoretical investigation of the dynamics of a droplet walking on a vibrating fluid bath under the influence of a harmonic potential. The walking droplet's horizontal motion is described by an integro-differential trajectory equation, which is found to admit steady orbital solutions. Predictions for the dependence of the orbital radius and frequency on the strength of the radial harmonic force field agree favorably with experimental data. The orbital quantization is rationalized through an analysis of the orbital solutions. The predicted dependence of the orbital stability on system parameters is compared with experimental data and the limitations of the model are discussed.
\end{abstract}

\maketitle
\section{Introduction}
There has been considerable recent interest in the dynamics of silicone oil droplets bouncing on the surface of a vibrating fluid bath \cite{Walker_Nature,JBAnnRev}. As discovered a decade ago in the laboratory of Couder and Fort, these droplets may move horizontally, or \lq walk,\rq\, across the fluid surface, propelled by the waves they generate at each bounce \cite{JFM_Suzie,Walker_Nature}. These walkers, comprising a bouncing droplet and an associated guiding wave field, exhibit behaviors reminiscent of quantum mechanical phenomena, including single-particle diffraction and interference \cite{Couder_Diffraction}, tunneling \cite{Eddi_Tunnel}, Zeeman-like splitting \cite{Eddi_PRL_2012}, orbital quantization in a rotating frame \cite{Fort_PNAS, Harris_JFM_2014}, and wave-like statistics in a confined geometry \cite{Harris_PRE_2013,Gilet2014}. The walking droplet system represents a hydrodynamic realization of the pilot-wave dynamics championed by de Broglie as an early model of quantum dynamics \cite{dB1,dB2}. The relationship between this hydrodynamic system and more modern realist models of quantum dynamics is explored elsewhere \cite{JBAnnRev,JBPhysToday}.

\begin{figure}
  \centerline{\includegraphics[scale=.42]{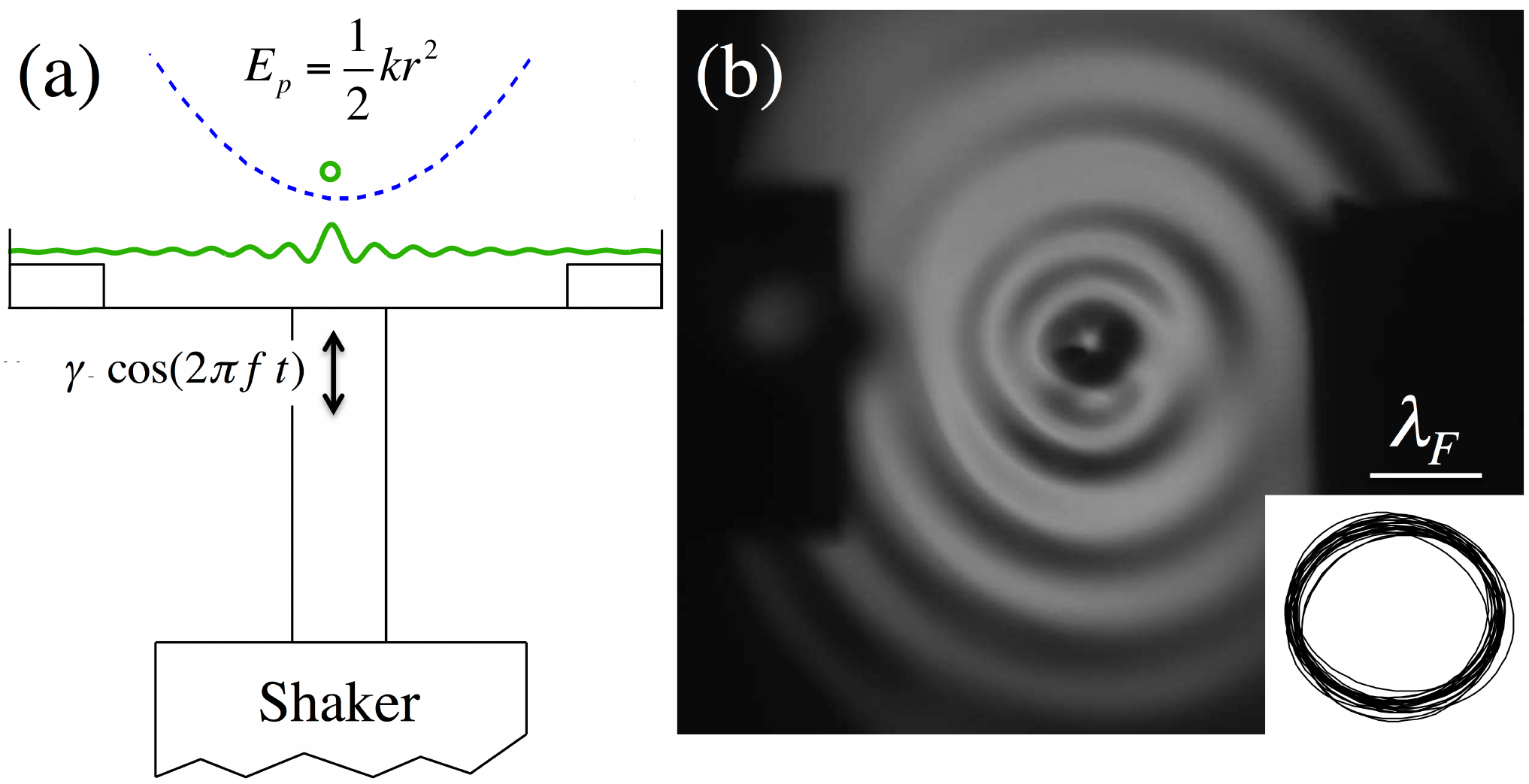}}
  \caption{(a) Schematic of the experimental setup, in which an oil droplet encapsulates a small amount of ferrofluid and is trapped in a harmonic potential $E_p = k\boldsymbol{r}^2/2$. The harmonic potential remains a good approximation up to distances of approximately $3\lambda_F \simeq 14$ mm (see \cite{Perrard_Nature_2014} for details). The fluid bath is driven vertically with acceleration $\gamma\cos(2\pi ft)$. (b) Top view of the walker and its associated wave field. The Inset shows a  characteristic circular trajectory. The scale bar $\lambda_F=4.75$ mm.}
\label{Schematics}
\end{figure}

	We consider here an experiment performed by Perrard {\it et al.} \cite{Perrard_Nature_2014}, in which the walking droplet moves in a two-dimensional harmonic potential. The experimental setup is shown in Fig.~\ref{Schematics}; the details have been presented elsewhere \cite{Perrard_Nature_2014}. The droplet encapsulates a small amount of ferrofluid and acquires a magnetic moment when placed in the spatially homogeneous magnetic field induced by two large Helmholtz coils. It is then attracted toward the symmetry axis of a cylindrical magnet suspended above the fluid bath. Provided the walker is not too far from the magnet's axis, a radially inward force is generated on the drop. The force $\bm{F} = -k\bm{x}$ increases linearly with distance from the magnet's axis, where $\bm{x}$ is the displacement from the origin, and the constant $k$ may be tuned by adjusting the vertical distance between the magnet and the fluid bath.

	Perrard {\it et al.} \cite{Perrard_Nature_2014} reported that the walker dynamics in a harmonic potential is sensitive to the {\it memory parameter}, as prescribed by the proximity to the Faraday threshold, which determines the longevity of the standing waves generated by the walker  \cite{Eddi_JFM_2011}. In the low-memory limit, in which the waves decay relatively quickly, the walker executes circular orbits whose radii decrease monotonically with increasing spring constant $k$. As the memory parameter is increased, the orbital radii become quantized. The authors also reported the existence of other periodic and quasiperiodic trajectories, such as the trefoil and lemniscate. The various trajectories were found to be quantized in both mean radius and angular momentum. In the high-memory limit, the walker exhibits a chaotic dynamics characterized by intermittent
transitions between a set of quasiperiodic trajectories~\cite{Perrard_PRL_2014}. Labousse \textit{et al.}~\cite{Labousse_NJP_2014} linked this complex 
dynamics to the self-organization of its wave field.\\	
	
The first theoretical model of the walker system, developed by Proti\`{e}re \textit{et al.}~\cite{JFM_Suzie}, captured 
certain features of the observed behavior, including a transition from bouncing to walking. The understanding of richer phenomena required the inclusion of memory effects~\cite{Eddi_JFM_2011}, in which the past bounces are encoded in the surface wave field. Through an analysis of the droplet impact and the resulting standing waves, Mol\'{a}\v{c}ek and Bush~\cite{Molacek_POF_2012,Molacek_JFM_1_2013,Molacek_JFM_2_2013}
derived a trajectory equation for the walker that includes both its vertical and horizontal dynamics. By averaging out the vertical dynamics, Oza {\it et al.}~\cite{Oza_JFM_1_2013} derived an integro-differential form for the 
horizontal motion referred to as the stroboscopic model. This theoretical framework provides a valuable platform for analytical investigations. For example, \color{black}the resulting equation was used to derive 
reduced trajectory equations appropriate in the limits of low-memory~\cite{Labousse_PRE1_2014} and weak horizontal 
acceleration~\cite{Boosht}.\\ 

Fort {\it et al.} \cite{Fort_PNAS} and subsequently Harris and Bush \cite{Harris_JFM_2014} examined droplets walking in a rotating frame. 
The walkers were found to execute circular inertial orbits provided the memory was sufficiently low. 
In the low-memory limit, the orbital radius decreased monotonically with the applied rotation rate.
As the memory was progressively increased, the circular orbits became quantized in radius. 
Fort {\it et al.} \cite{Fort_PNAS} presented numerical simulations that captured the emergence of orbital quantization with 
increasing memory. This quantization was rationalized in terms of a theoretical model based on considering 
the composite effect of wave sources on a circle in the high-memory limit. Oza {\it et al.} \cite{Oza_JFM_2_2013,ExoticOrb} augmented the
stroboscopic model~\cite{Oza_JFM_1_2013} through inclusion of the Coriolis force, $\bm{F}_{\mathrm{Cor.}}=-2m\boldsymbol{\Omega}\times\dot{\bm{x}}_p$, in order to rationalize the orbital 
stability thresholds and complex dynamics reported in the experimental study of Harris and Bush \cite{Harris_JFM_2014}.
We adopt here a similar methodology, based on the stroboscopic model, in order to rationalize the orbital 
quantization of circular orbits arising in a harmonic potential, as reported in the experiments of Perrard {\it et al.} \cite{Perrard_Nature_2014}.

	The paper is organized as follows. We first present the integro-differential trajectory equation for the walker in the presence of an external confining potential and show that it admits orbital solutions. We compare our model with the existing experimental data obtained for a harmonic potential~\cite{Perrard_Nature_2014}. We restrict our investigation to the circular orbits. We then analyze the linear stability of the orbital solutions and compare our results to laboratory experiments of walkers in a harmonic well. We use the stability analysis to rationalize the emergence of quantization of circular orbits and discuss the discrepancies between the theoretical predictions and experimental data. Finally, we link the orbital instabilities to wave modes excited by the walker.  We conclude by discussing future directions.

\section{Existence of quantized orbits}\label{Sec:TrajEqn}
\subsection{Trajectory equation}
Consider a drop of mass $m$ and undeformed radius $R_d$ walking on the surface of a vertically vibrating fluid bath of density $\rho$, surface tension $\sigma$, kinematic viscosity $\nu$, mean depth $H$ and vertical acceleration $\gamma\cos(2\pi ft)$. We restrict our attention to the regime $\gamma < \gamma_F$, $\gamma_F$ being the Faraday instability threshold \cite{FaradayOriginal,Benjamin_Ursell,Douady_JFM,EPLDouadyFauve,Tuckerman_JFM_2009}, below which the fluid surface would remain flat in the absence of disturbances. Theoretical treatments have been developed to rationalize the drop's bouncing dynamics \cite{JFM_Suzie,PRELifetimedroplet,Terwagne_Phys_2008,NJOPRoller, Molacek_POF_2012,Molacek_JFM_1_2013,Oistein_POF_2013,Hubert_2014,Milewski_2015,Blanchette_2016}. We restrict our study to the particular case in which the drop is in a perfectly period-doubled bouncing state, as is typically the case in the walking regime \cite{Molacek_JFM_2_2013}. The drop's bouncing period $T_F = 2/f$ is then commensurate with its subharmonic Faraday wave field \cite{JFM_Suzie,Benjamin_Ursell}. Assuming that the drop hits the bath with a constant phase relative to the vibrational forcing, we may consider 
the simplified strobed dynamics for the droplet's horizontal motion \cite{Oza_JFM_1_2013}. 

Let $\bm{x}_p(t)=(x_p(t),y_p(t))$ be the horizontal position of the walker at time $t$. During each impact, the walker experiences a propulsive force proportional to the local slope of the fluid interface and a drag force opposing its motion. Time averaging these forces on the drop over the bouncing period $T_F$ yields the equation of motion \cite{Molacek_JFM_2_2013}
\begin{align}
\label{trajeqn}
m\ddot{\bm{x}}_p+D\dot{\bm{x}}_p= \bm{F}-mg\bm{\nabla} h(\bm{x}_p,t) ,
\end{align}
where $h(\bm{x},t)$ is the height of the fluid interface and $\bm{F}$ is an arbitrary external force on the drop. The time-averaged drag coefficient $D$ has the form \cite{Molacek_JFM_2_2013} $D = Cmg\sqrt{\frac{\rho R_d}{\sigma}}+6\pi\mu_a R_d\left(1+\frac{\rho_agR_d}{12\mu_af}\right)$, where $\mu_a$ and $\rho_a$ are the dynamic viscosity and density of air, respectively, and the coefficient $C = 0.17$ is inferred from the drop's tangential coefficient of restitution. The first term in $D$ accounts for the direct transfer of momentum from drop to bath during impact and the second accounts for air drag. 

The wave field resulting from the drop's prior impacts may be written as~\cite{Eddi_JFM_2011,Molacek_JFM_2_2013}
\begin{align}
\label{hsum}
h(\bm{x},t) = \sum_{n=-\infty}^{\lfloor t/T_F\rfloor} AJ_0\left(k_F\left|\bm{x}-\bm{x}_p(nT_F)\right|\right)\mathrm{e}^{-\frac{t-nT_F}{M_e T_F}},
\end{align}
where the memory parameter $M_e$ is given by $M_e(\gamma)= \frac{T_d}{T_F\left(1-\gamma/\gamma_F\right)}.$ The hydrodynamic analysis of Mol\'{a}\v{c}ek and Bush~\cite{Molacek_JFM_1_2013} demonstrated that the wave amplitude may be expressed as $A= \frac{1}{2}\sqrt{\frac{\nu}{T_F}}\frac{k_F^3}{3k_F^2\sigma+\rho g}mgT_F\sin\Phi$. The Faraday wavenumber $k_F$ is defined through the standard water-wave dispersion relation $(\pi f)^2 = \left(gk_F+\sigma k_F^3/\rho\right)\tanh(k_FH)$. Here $\Phi$ is the mean phase of the wave during the contact time and $T_d\approx 0.0182$ s the viscous decay time of the waves in the absence of forcing for $\nu = 20$ cS \cite{Molacek_JFM_2_2013}. We note that the phase $\sin\Phi$ may be deduced from the experimentally observed free walking speed \cite{Oza_JFM_1_2013}. The memory parameter $M_e$ increases with the forcing acceleration $\gamma$ and determines the extent to which the walker is influenced by its past \cite{Eddi_JFM_2011}. Indeed, the dominant contribution to the wave field (\ref{hsum}) comes from the drop's $n\sim O(M_e)$ prior bounces. 

Provided the time scale of horizontal motion $T_H \sim \lambda_F/|\dot{\bm{x}}_p|$ is much greater than the bouncing period $T_F$, as is the case for walkers, we may approximate the sum in Eq.~(\ref{hsum}) by an integral \cite{Oza_JFM_1_2013}
\begin{align}
h(\bm{x},t) = \frac{A}{T_F}\int\limits_{-\infty}^t \mathrm{d}T\; J_0\left(k_F\left|\bm{x}-\bm{x}_p(T)\right|\right)\mathrm{e}^{-\frac{t-T}{M_e T_F}}.
\end{align}
This continuous approximation will allow us to compute analytical time-dependent behaviors by considering a perturbative approach. 
It therefore provides a framework for investigating a pilot-wave dynamics closely related to the walker's dynamics. The limits of validity of the continuous approximation will be discussed in what follows.

	We introduce the dimensionless variables $\hat{\bm{x}} = k_F\bm{x}$ and $\hat{t} = t/(T_FM_e)$. The external central force $\bm{F}$ may be expressed in dimensionless form as $\bm{\mathcal{F}}=(k_F M_e T_F /D)\bm{F}$.  The dimensionless trajectory equation~(\ref{trajeqn}) thus assumes the form
\begin{equation}
\begin{array}{ll}
\kappa\hat{\bm{x}}^{\prime\prime}_p+\hat{\bm{x}}^{\prime}_p&=\bm{\mathcal{F}}+ \\
&\displaystyle \beta\int\limits_{-\infty}^{\hat{t}}\mathrm{d}\hat{T}\;\bm{u}_{t,T} \;J_1\left(\left|\hat{\bm{x}}_p(\hat{t})-\hat{\bm{x}}_p(\hat{T})\right|\right)\mathrm{e}^{-(\hat{t}-\hat{T})}
\end{array}
\label{NDIDEqnGen}
\end{equation}
where primes denote differentiation with respect to $\hat{t}$, and $\kappa = m/(T_FM_e D)$, $\beta = mgAk_F^2 T_FM_e^2/D$ are the dimensionless mass and wave force coefficient, respectively and $\bm{u}_{t,T}$ denotes the unit vector pointing from $\hat{\bm{x}}_p(\hat{T})$ to $\hat{\bm{x}}_p(\hat{t})$. 

We now seek orbital solutions to the trajectory equation and so substitute $\hat{\bm{x}}_p(\hat{t})=r_0(\cos\omega\hat{t},\sin\omega\hat{t})$ into Eq.~(\ref{NDIDEqnGen}), where $\omega$ and $r_0$ are the dimensionless angular frequency and orbital radius, respectively. Dropping all carets, we obtain in polar coordinates $(r,\theta)$ the system of algebraic equations
\begin{align}
\label{NDimOrb}
-\kappa r_0\omega^2&=\beta\int_0^{\infty}J_1\left(2r_0\sin\frac{\omega z}{2}\right)\sin\frac{\omega z}{2}\mathrm{e}^{-z}\,\mathrm{d}z+\mathcal{F}_{r},\nonumber \\
r_0\omega &=\beta\int_0^{\infty}J_1\left(2r_0\sin\frac{\omega z}{2}\right)\cos\frac{\omega z}{2}\mathrm{e}^{-z}\,\mathrm{d}z+\mathcal{F}_{\theta}.
\end{align}

\subsection{Orbital solutions in a harmonic potential}
\begin{figure}
 \centering
    \includegraphics[width=\columnwidth]{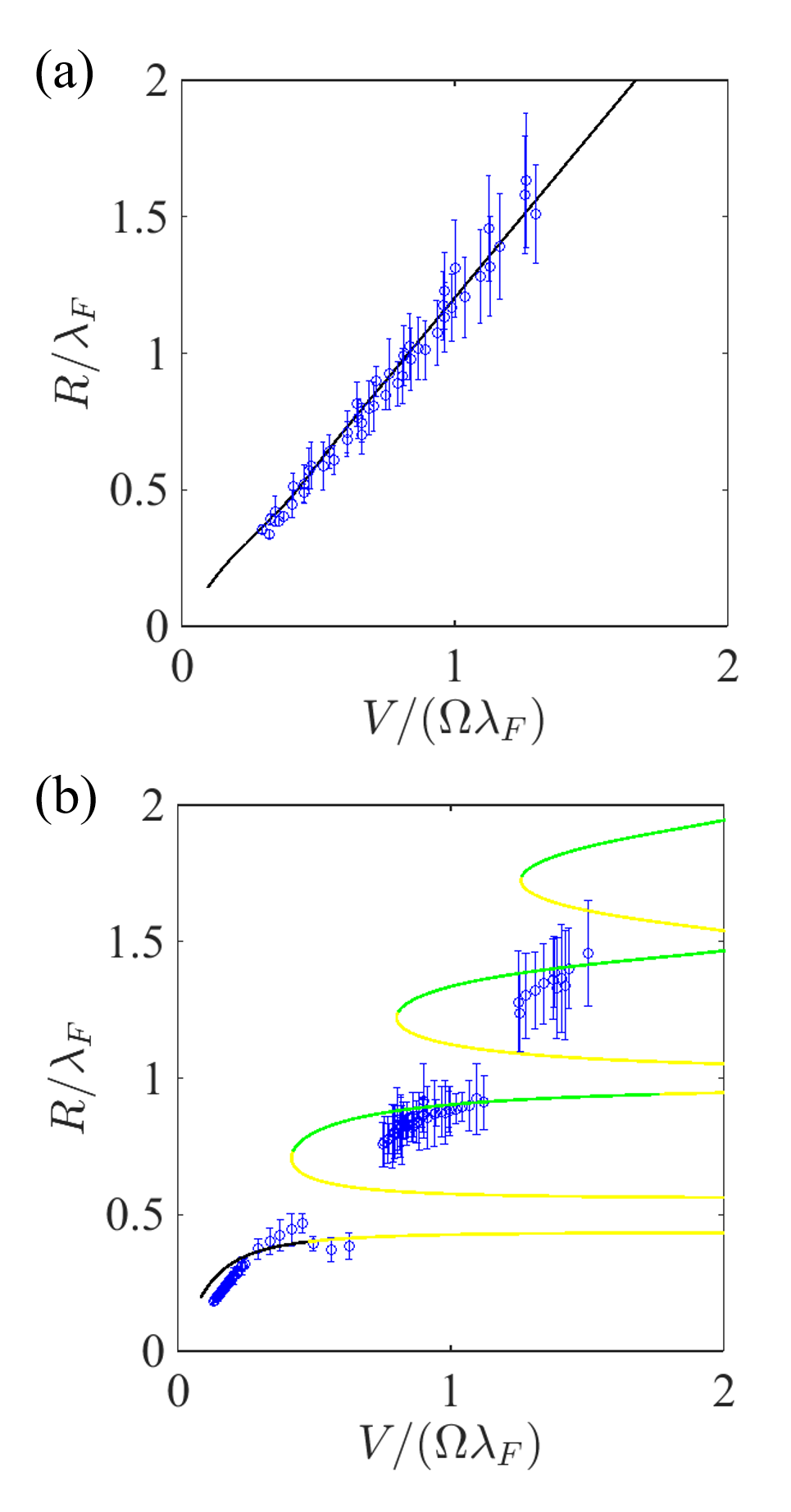}
 \caption{Evolution of the orbital radius $R/\lambda_F=r_0/(2\pi)$ with the potential width $\Lambda=V/(\Omega \lambda_F)$, where $\Omega=\sqrt{k/m}$. (a) At low memory $\gamma/\gamma_F \simeq 0.92 $, the radius increases linearly with the potential width. (b) In the high memory regime 
 $\gamma/\gamma_F = 0.979$, the orbital radii converge to regularly spaced plateaus. The curves indicate the theoretical predictions based on Eq.~\ref{NDIDEqnMot} and the colors refer to the linear stability analysis of orbital solutions described in Sec.~\ref{Sec:OrbStab}. Black denotes stable orbits, green denotes unstable orbits that destabilize via an oscillatory instability, and yellow indicates the coexistence of oscillatory and non oscillatory unstable modes. The lower and upper horizontal cross-cuts evident in Fig.~\ref{Stability} correspond to the two data sets shown here in (a) and (b), respectively.}
\label{Orbitals}
\end{figure}

For a harmonic potential, we have $\bm{F} = -k\bm{x}_p$ or equivalently $\bm{\mathcal{F}}=-\xi \bm{\hat{x}}_p$, $\xi = kT_FM_e/D$ being the dimensionless strength of the harmonic potential. Equation~(\ref{NDimOrb}) thus takes the form
\begin{align}
\label{NDIDEqnMot}
-\kappa r_0\omega^2&=\beta\int_0^{\infty}J_1\left(2r_0\sin\frac{\omega z}{2}\right)\sin\frac{\omega z}{2}\mathrm{e}^{-z}\,\mathrm{d}z-\xi r_0,\nonumber \\
r_0\omega &=\beta\int_0^{\infty}J_1\left(2r_0\sin\frac{\omega z}{2}\right)\cos\frac{\omega z}{2}\mathrm{e}^{-z}\,\mathrm{d}z.
\end{align} 
Given the experimental parameters that determine $\kappa$, $\beta$ and $\xi$, these equations can be solved numerically using computational software (MATLAB), which yields the orbital radius $r_0$ and frequency $\omega$ of the circular orbit.  The dependence of the orbital radius on the dimensionless potential width $\Lambda=V/(\lambda_F\sqrt{k/m})$ is shown in Fig.~\ref{Orbitals} for two different values of forcing acceleration $\gamma$, $V$ being the walker's time-averaged horizontal speed. At low memory $\gamma/\gamma_F \simeq 0.92$ [Fig.~\ref{Orbitals}(a)], the orbital radius increases monotonically with the potential width $\Lambda$. The linear dependence may be understood  from the balance of the attractive force and the centripetal acceleration. The slope exceeding one is a signature of the wave-induced added mass, which may be expressed in terms of a hydrodynamic boost factor~\cite{Boosht}. At higher memory (fig.~\ref{Orbitals}b), the orbital radius exhibits a non monotonic dependence on the potential width $\Lambda$, leading to pronounced plateaus with forms consistent with those reported by Perrard \textit{et al.}~\cite{Perrard_Nature_2014}. In the following section we will see  that the yellow branches of the solution curves in [Fig.~\ref{Orbitals}(b)] correspond to unstable solutions.

\subsection{Orbital solutions for any central force in the limit $M_e \gg 1$}
For the sake of generality, we now examine the condition for the existence of quantized circular orbits for any axisymmetric confining potential. Let us start with Eq.~(\ref{NDimOrb}), which represents the radial and tangential balance of forces. As the memory increases, the radial terms of Eq.~(\ref{NDimOrb}) scale as
\begin{equation}
\left\{
\label{EqnHighM}
\begin{array}{ll}   
-\kappa r_0\omega^2 \sim \mathit{O}\left(M_e \right)\\
\mathcal{F} \sim \mathit{O}\left(M_e \right)\\
\beta\int_0^{\infty}J_1\left(2r_0\sin\frac{\omega z}{2}\right)\sin\frac{\omega z}{2}\mathrm{e}^{-z}\,\mathrm{d}z\sim \mathit{O}\left(M_e^2 \right)
 \end{array}
 \right.
\end{equation}
\noindent and therefore act on different time scales. At high memory, the long-time-scale terms dominate, yielding 
\begin{equation}
\int_0^{\infty}J_1\left(2r_0\sin\frac{\omega z}{2}\right)\sin\frac{\omega z}{2}\mathrm{e}^{-z}\,\mathrm{d}z=\mathit{O}\left(\dfrac{1}{M_e} \right).
\label{Longts}
\end{equation}
As in the case of inertial orbits~\cite{Oza_JFM_2_2013}, Eq.~(\ref{Longts}) admits a set of orbital solutions
\begin{equation}
r_0^{(n)}=\jmath_{0,n}+\mathit{O}\left(\dfrac{1}{M_e} \right)
\end{equation}
where $\jmath_{0,n}$ is the $n$-th zero of the Bessel function $J_0$. 
These orbital solutions correspond to the plateaus observed in Fig.~\ref{Orbitals}(b). Provided $\mathcal{F}$ is not singular in $r_0^{(n)}$, a set of quantized orbital solutions will arise at high memory. The $\mathit{O}\left(1/M_e \right)$ corrections depend on the form of the potential and will determine the exact value of $r_0^{(n)}$ but will not affect the existence of solutions. We thus turn to the stability of these orbital solutions using the continuous approximation.
\section{Linear stability of orbital solutions}\label{Sec:OrbStab}
\subsection{General case}
We perform a linear stability analysis of circular orbital solutions in the presence of an arbitrary radial force $\mathcal{F}(r,\xi)$, where $r = |\hat{\bm{x}}|$ and $\xi$ is a parameter that controls the strength of the force. For the harmonic potential of interest, $\mathcal{F}(r,\xi) = -\xi r$. We linearize the trajectory equation (\ref{NDIDEqnGen}) around the orbital solution defined by Eq.~(\ref{NDimOrb}), substituting $r=r_0+\epsilon r_1(t)$ and $\theta=\omega t+\epsilon\theta_1(t)$ into Eq.~(\ref{NDIDEqnGen}) and retaining terms to leading order in $\epsilon$. We note that the presence of the convolution product in the linearized equations of motion
indicates the presence of long-range temporal correlations in the dynamics that complicate the stability analysis. \\

	We take the Laplace transform ${\cal L}$ of the linearized equations and obtain a system of algebraic equations for $R(s)={\cal L}[r_1]$ and $\Theta(s)={\cal L}[\theta_1]$:
\begin{eqnarray}
\begin{pmatrix} A(s) & -B(s) \\ C(s) & D(s)\end{pmatrix}\begin{pmatrix} R(s) \\ r_0\Theta(s)\end{pmatrix} = \begin{pmatrix} c_r \\ r_0c_\theta\end{pmatrix}
\label{EqnLin}
\end{eqnarray}
where
\begin{align}
\label{ABCD1}
A(s)&=\kappa s^2+s-\kappa\omega^2-\left.\pdiff{{\cal F}}{r}\right|_{r_0}-\beta\left(\int_0^\infty\left[f(t)\cos^2\frac{\omega t}{2}\right.\right.\nonumber \\
&\phantom{=}\left.\left.+g(t)\sin^2\frac{\omega t}{2}\right]\,\mathrm{d}t+\mathcal{L}\left[g(t)\sin^2\frac{\omega t}{2}-f(t)\cos^2\frac{\omega t}{2}\right]\right),\nonumber \\
B(s) &= 2\kappa\omega s-\left(\kappa\omega+\frac{{\cal F}(r_0,\xi)}{r_0\omega}\right)\nonumber \\
&\phantom{=}-\frac{\beta}{2}\mathcal{L}\left[\left(f(t)+g(t)\right)\sin\omega t\right],\nonumber \\
C(s) &= 2\kappa\omega s+2\omega+\kappa\omega+\frac{{\cal F}(r_0,\xi)}{r_0\omega}\nonumber \\
&\phantom{=}-\frac{\beta}{2}\mathcal{L}\left[\left(f(t)+g(t)\right)\sin\omega t\right],\nonumber \\
D(s)&=\kappa s^2+s-1-\beta\mathcal{L}\left[f(t)\sin^2\frac{\omega t}{2}-g(t)\cos^2\frac{\omega t}{2}\right].
\end{align}
Here $f(t)=\frac{J_1\left(2r_0\sin\frac{\omega t}{2}\right)}{2r_0\sin\frac{\omega t}{2}}\mathrm{e}^{-t}$, $g(t) = J_1^{\prime}\left(2r_0\sin\frac{\omega t}{2}\right)\mathrm{e}^{-t}$, and we have used Eq.~(\ref{NDimOrb}) to simplify some of the integrals. The constants $c_r$ and $c_\theta$ are defined through the initial conditions by $r_1(0)=c_r/\kappa$ and $\theta_1(0)=c_{\theta}/\kappa$ \cite{Oza_JFM_2_2013}. The poles of the linearized equation (\ref{EqnLin}) are the roots of the function $G(s;r_0)\equiv A(s)D(s)+B(s)C(s)$. If all of the roots satisfy $\mathrm{Re}(s) < 0$, the orbital solution of radius $r_0$ is stable to perturbations, while a single root in the right half-plane is sufficient for instability. To assess the stability of an arbitrary orbital solution, we find the roots of $G(s;r_0)$ numerically. Since $G(s;r_0)$ has poles at $s=-1 + in\omega$ for integers $n$, we instead find the roots of the function $\tilde{G}(s;r_0)=(1-\mathrm{e}^{-2\pi (s+1)/|\omega|})G(s;r_0)$, which is an entire function of $s$. We find the roots of $\tilde{G}$ numerically by implementing the integral method of Delves and Lyness~\cite{Delves_1966}. We took the precaution of benchmarking this root tracking method in order to assess the precision of our numerical method.

\subsection{Stability diagram for circular orbits in a harmonic potential} 
\begin{figure*}
 \centering
    \includegraphics[width=2\columnwidth]{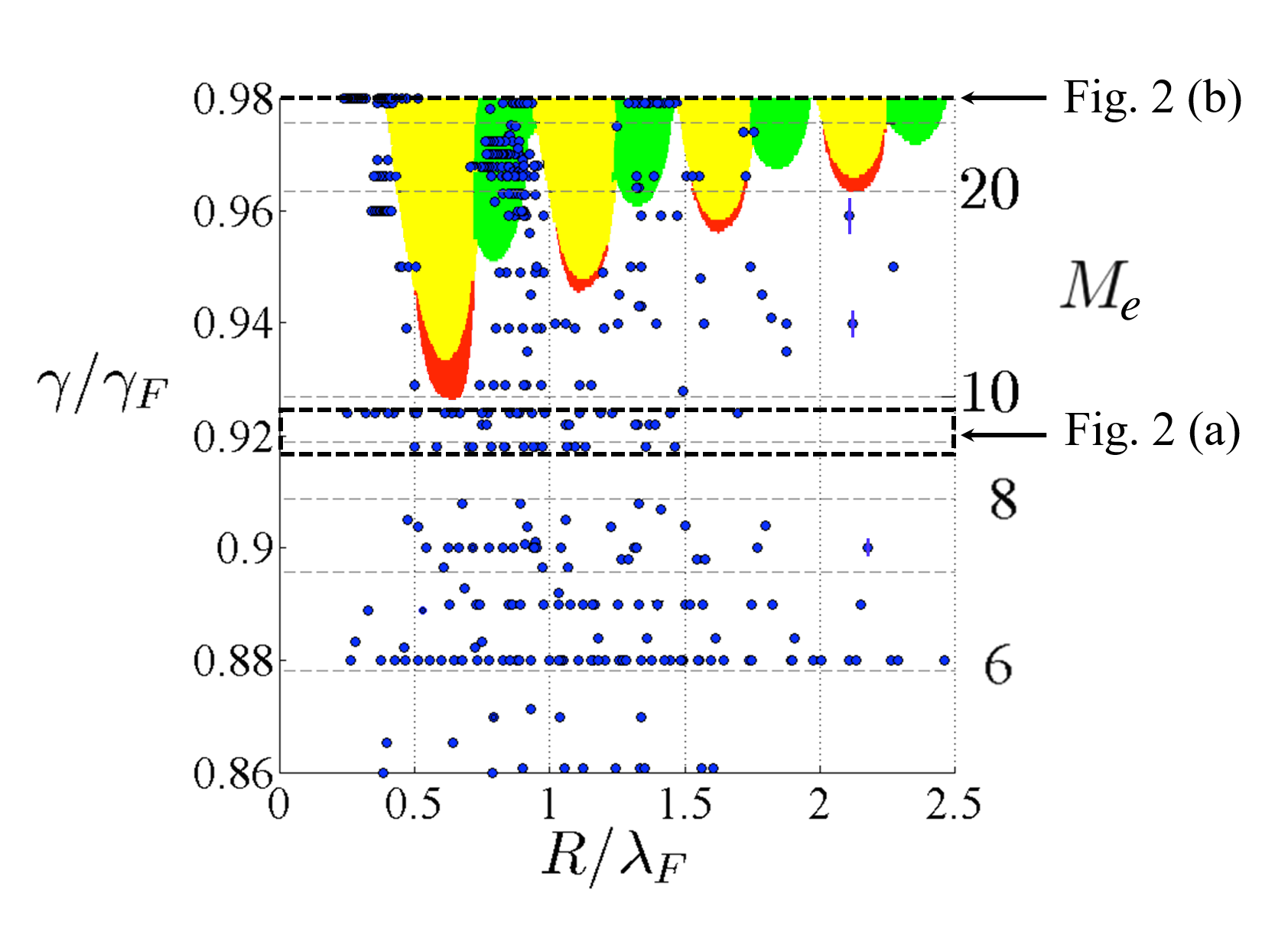}
 \caption{Stability of orbital solutions in a harmonic potential. Colored regions indicate the linearly unstable parameter regime, as predicted theoretically. Red indicates solutions that destabilize via a non oscillatory instability, green indicates solutions that destabilize via an oscillatory instability and yellow indicates solutions with coexisting oscillatory and non-oscillatory unstable modes. Blue solid circles are the experimental data from Perrard {\it et al.}~\cite{Perrard_Nature_2014}. The horizontal cross cuts correspond to the data reported in Figs 2(a) and 2(b). Characteristic error bars are shown.}
\label{Stability}
\end{figure*}

We performed the stability analysis for the specific case of a harmonic potential; $\mathcal{F}(r,\xi) = -\xi r$;
In Fig.~\ref{Stability}, we present the results of the orbital stability analysis for a drop of radius $R_d = 0.37$ mm and phase $\sin\Phi = 0.18$ walking on a fluid bath of viscosity $\nu = 20$ cS and forcing frequency $f = 80$ Hz, the parameters being inferred from the experiments of Perrard {\it et al.} \cite{Perrard_Nature_2014}. We note that multiple orbital solutions may exist for a given value of the spring constant $k$, but that the orbital solution is uniquely determined by the orbital radius $r_0$ and forcing acceleration $\gamma/\gamma_F$, and so plot the orbital stability properties on the $(R/\lambda_F,\gamma/\gamma_F)$ plane with $R/\lambda_F=r_0/(2\pi)$. The stability of a given orbital solution is determined by the roots of $\tilde{G}(s;r_0)$, denoted by $s_*$, and indicated by the following color code in Figs~\ref{Orbitals} and~\ref{Stability}.  Black in Fig.~\ref{Orbitals} and white in Fig.~\ref{Stability} denote orbital solutions that are stable to perturbations ($\mathrm{Re}(s_*) < 0$). Green denotes solutions that destabilize via an oscillatory instability ($\mathrm{Re}(s_*) > 0,\mathrm{Im}(s_*) \neq 0$). Red refers to unstable cases with a non oscillatory mechanism ($\mathrm{Re}(s_*) > 0,\mathrm{Im}(s_*) = 0$). Finally, yellow indicates the coexistence of oscillatory and non-oscillatory unstable modes.
Figure~\ref{Stability} shows adequate agreement between the predictions of our stability analysis and the experimental results of Perrard {\it et al.}~\cite{Perrard_Nature_2014}, in the sense that none of the experimental data points indicating stable circular orbits 
fall within the red or yellow regions. 

The principal discrepancy between our theoretical predications and the observed orbital stability
is evident in the data points arising at high memory ($\gamma/\gamma_F > 0.95$) within the green regions, where the linear 
theory predicts an oscillatory instability. 
In the investigation of quantization of inertial orbits \cite{Oza_JFM_2_2013}, stable orbits were also observed in a regime predicted to be unstable via linear stability analysis. There the orbits were wobbling circular 
orbits  \cite{Harris_JFM_2014}, presumed to have been stabilized by nonlinear effects.  Here the observed orbits 
were not wobbling significantly, though there are practical difficulties in distinguishing stable circular orbits 
from small-amplitude wobbling orbits. We believe the mismatch arising at high memory to be due to shortcomings of
our theoretical model, specifically, the stroboscopic approximation.

In our theoretical treatment, we make a number of simplifying assumptions that could explain the discrepancy between 
theory and experiment. First, the stroboscopic approximation (\ref{NDIDEqnGen}) rests on the assumption of perfect 
synchronization between the drop and wave, a synchronization that may break down in the high-memory regime, 
where asynchronous chaotic walking states may arise \cite{Oistein_POF_2013}.
Second, we assume the phase $\Phi$ to be a constant, whereas it is known to vary weakly with forcing acceleration 
\cite{Molacek_JFM_2_2013} and is also expected to depend on the local wave amplitude. 
Finally, it is known that a differential equation and its discretized form may possess different instabilities~\cite{Allaire_book}. In future work we plan to examine the stability of orbital solutions with a discretized version of the trajectory equation (\ref{NDIDEqnGen}), thereby assessing the relative merits of the continuous and discrete approaches.

\subsection{Mode decomposition of orbital instabilities}

\begin{figure*}
 \centering
    \includegraphics[width=1.5\columnwidth]{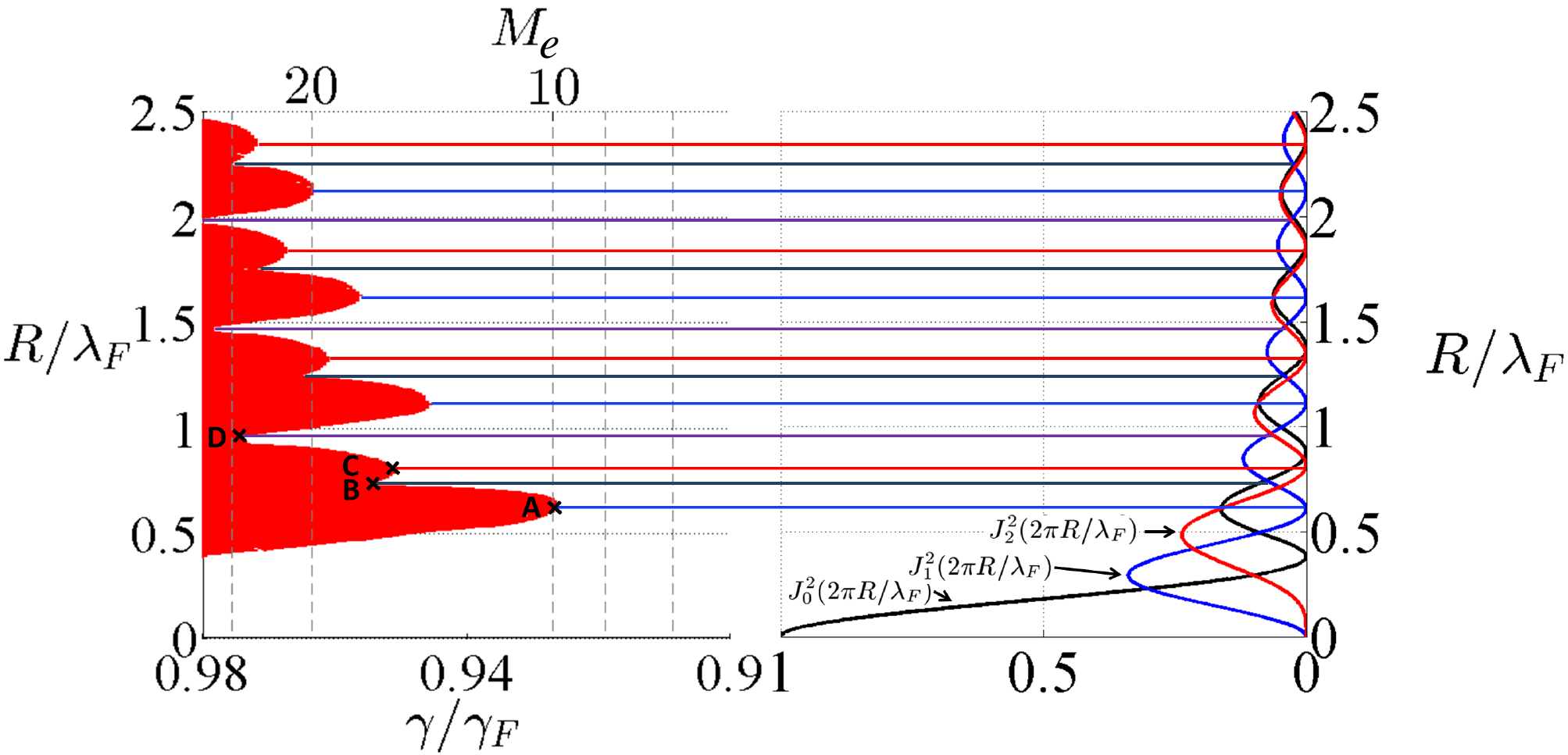}
 \caption{Striking link between the points of the stability diagram (left) and the Bessel modes (right). The black curve corresponds to $J_0^2(2\pi R/\lambda_F)$, the blue curve to $J_1^2(2\pi R/\lambda_F)$, and the red curve to $J_2^2(2\pi R/\lambda_F)$. }
\label{Stabilitywave}
\end{figure*}

We now infer a connection between the nature of the radial force on an orbiting walker and the results of our stability analysis in Fig.~\ref{Stability}. Using Graf's addition theorem, the radial force balance in Eq.~\eqref{NDIDEqnMot} may be written as
\begin{align}
-\kappa r_0\omega^2=-\beta\left[ \pdiff{}{r}\sum_{p=0}^{\infty}\frac{(2-\delta_{n,0})}{(1+(p\omega)^2)}J_p(r)J_p(r_0)\right]_{r=r_0}-\xi r_0.
\end{align}
That is, the orbiting walker experiences a potential energy comprised of the weighted sum of modes $J_p^2(r_0)$. Plots of these modes for $p= 0$, 1 and 2 are shown in Fig.~\ref{Stabilitywave}, along with the orbital stability diagram from Fig.~\ref{Stability}. Note that the red instability regions originate (point A) near the zeros of $J_1^2(r_0)$ and the green ones (point C) near the zeros of $J_2^2(r_0)$. This seems to suggest that the primary and secondary orbital instabilities occur for orbits that receive little energy from the $p=1$ and $2$ modes, respectively. We also observe that the intersection of neighboring instability regions may be related to the points at which these energetic modes assume the same value. Indeed, point B in Fig.~\ref{Stabilitywave} corresponds to the intersection between the $p=0$ and $1$ modes, and point D to that between the $p=1$ and $2$ modes. Establishing a precise connection between the modes $J_p^2(r)$ and the walker's orbital stability properties is beyond the scope of this paper. For the time being, we simply hypothesize that the small number of modes involved in the orbital instability is directly connected to the low-dimensional chaos observed in laboratory experiments of walker dynamics in a harmonic potential \cite{Perrard_PRL_2014} and a rotating frame \cite{Harris_JFM_2014}.

\section{Conclusion}\label{Sec:Conc}
We have presented a theoretical investigation into the orbital dynamics of a walking droplet subject to a spring force, $\bm{F} = -k\bm{x}$. The integro-differential trajectory equation~\eqref{NDIDEqnGen} for the walker's horizontal motion was shown to have orbital solutions, in which a walker follows a circular trajectory of radius $r_0$ with a fixed angular frequency $\omega$. The predicted dependence of $r_0$ on the dimensionless potential width $\Lambda$ adequately matches the experimental data of Perrard {\it et al.} \cite{Perrard_Nature_2014}, as shown in Fig.~\ref{Orbitals}. This analysis thus serves to rationalize the quantization of orbital radius $r_0$, as observed in laboratory experiments \cite{Perrard_Nature_2014}. 

The results of the stability analysis are summarized in Fig.~\ref{Stability}, which shows the dependence of the walker's orbital stability characteristics on the orbital radius $r_0$ and vibrational forcing $\gamma/\gamma_F$. The match between our theoretical predictions and the experimental data is encouraging, although  a number of circular orbits were observed within the theoretically predicted green instability regions at high memory. Although the continuous equation gives an adequate framework to deal with the integro-differential equation of motion, some discrepancies still need to be explored and understood. Possible sources of this discrepancy have been discussed. 

While the linear stability analysis presented herein helps to delineate the parameter regimes in which circular motion is unstable, it does not provide a rationale for any of the other reported forms of stable motion (such as lemniscates and trifoliums) or for the complex walker dynamics arising within the unstable regions. These regions appear to be characterized by a self-organization mechanism between the drop trajectory and its associated wave field~\cite{Labousse_NJP_2014}. Consequently, in the high memory limit, an ordered chaos in the walker dynamics underlies the observed multi-modal statistical behavior \cite{Harris_JFM_2014,Gilet2014,ExoticOrb,Perrard_Nature_2014,Perrard_PRL_2014}. 
Much remains to be done in terms of rationalizing the connection between the dynamics and statistics in the high-memory limit.
\acknowledgements
The authors thank Y. Couder and E. Fort for encouraging this work and for useful discussions.  This research was supported by the French Agence Nationale de la Recherche, through the project \lq\lq ANR Freeflow,\rq\rq\,LABEX WIFI (Laboratory of Excellence ANR-10-LABX-24) within the French Program “Investments for the Future” under reference ANR-10- IDEX-0001-02 PSL and the AXA Research Fund. A.U.O. acknowledges the support of the NSF Mathematical Sciences Postdoctoral Fellowship. J.W.M.B gratefully acknowledges continuing support from the NSF through Grant CMMI-1333242.

\end{document}